# Strongly suppressed 1/f noise and enhanced magnetoresistance in epitaxial Fe-V/MgO/Fe magnetic tunnel junctions

D.Herranz[1], F. Bonell[2], A.Gomez-Ibarlucea[1], S. Andrieu[2], F. Montaigne[2], R.Villar[1], C. Tiusan[2], and F.G.Aliev[1]*

(1) Dpto. Fisica Materia Condensada, CIII, Universidad Autonoma de Madrid, 28049, Cantoblanco, Madrid Spain

(2) Institut Jean Lamour, CNRS-Nancy University, Bd. des Aiguillettes, B.P. 239, 54506 Vandouvre-lès-Nancy Cedex, France

Alloying Fe electrodes with V, through reduced FeV/MgO interface mismatch in epitaxial magnetic tunnel junctions with MgO barriers, notably suppresses both nonmagnetic (parallel) and magnetic (antiparallel) state 1/f noise and enhances tunnelling magnetoresistance (TMR). A comparative study of the room temperature electron transport and low frequency noise in $Fe_{1-x}V_x$/MgO/Fe and Fe/MgO/$Fe_{1-x}V_x$ MTJs with $0 \leq x \leq 0.25$ reveals that V doping of the bottom electrode for $x < 0.1$ reduces in nearly 2 orders of magnitude the normalized nonmagnetic and magnetic 1/f noise. We attribute the enhanced TMR and suppressed 1/f noise to strongly reduced misfit and dislocation density.

(*) corresponding author: farkhad.aliev@uam.es

Magnetic tunnel junctions (MTJs) with Fe/MgO/Fe structure are model systems where prediction [1,2] has been followed by discovery of giant TMR exceeding 100% at room temperature (RT) [3-8]. These findings have strongly intensified studies of the spin dependent coherent tunnelling both from fundamental and applied points of view. It is well known that there is a 3.9% lattice mismatch between Fe and MgO which induces stress within the MgO barrier. The strain is partially relaxed during the MgO growth via interfacial dislocations [6,7,9]. These may be partially responsible for reduced TMR below 1000% predicted theoretically [1,2] and should also determine substantially the defect related 1/f noise [10,11].

A previous study, focused on MTJs with relatively thick (2.5 nm MgO) barriers [12], aimed at employing some Fe alloy as a ferromagnetic electrode in order to reduce the interface lattice mismatch. It was shown that substitution of Fe with V decreases magnetic damping [13] and increases the lattice parameter of the electrode and therefore reduces the misfit [12]. It is important, however, to verify the implementation of a similar strategy in MTJs with thinner barriers and smaller resistance by area (RA) products, which are more relevant for applications. This type of interface engineering may be expected to enhance the low frequency electronic stability of MTJs via suppressed defect related noise. Indeed, the quality of the metal-insulator interface in MTJs determines the 1/f noise, mainly through defect induced charge trapping /de-trapping [14, 15].

Here we report on the detailed study of RT magnetoresistance and 1/f noise, in both parallel (P) state and antiparallel (AP) state, in fully epitaxial $Fe_{1-x}V_x$/MgO/Fe and Fe/MgO/$Fe_{1-x}V_x$ MTJs with a 9.5±0.5 ML thick MgO barrier. We have observed an increase of about 10 % of TMR with a relatively small V doping of the bottom electrode ($x$ about 0.1), and a remarkable reduction, in nearly two orders of magnitude, of the normalized 1/f noise (Hooge factor). Even a stronger noise reduction is observed in the AP state in the conditions of a substantially enhanced TMR.

Fe-V(50 nm)/MgO(2 nm)/Fe(18 nm)/Co(20nm)/Au(20 nm) and Fe(50 nm)/MgO(2 nm)/Fe-V(18 nm)/Co(20nm)/Au(20 nm) multilayers were grown by molecular beam epitaxy on MgO(001) substrates. Before the stacking deposition, substrates were outgassed at 875 K and a 7.5 nm thick MgO layer was grown in order to prevent the diffusion of residual carbon through the bottom electrode and its segregation at the interface with MgO [16]. Fe-V alloys were obtained by Fe and V coevaporation. The V concentration was checked after growth by X-ray photoelectron spectroscopy. The barrier thickness was controlled by RHEED intensity oscillations. The MTJs were then patterned by UV photolithography and Ar etching to lateral dimension ranged from 10 µm to 50 µm. Additional preparation details can be found in Ref [12].

A total of 61 MTJs samples were studied, with 8 on average for each V concentration. The TMR and 1/f noise have been studied at RT using a four-probe method. The dynamic conductance and TMR were measured using a current modulation superimposed on the dc current. The voltage noise power was studied in the frequency range between f=2-1600Hz using a cross correlation technique. We used the phenomenological Hooge factor (α) in order to compare the voltage noise power ($S_V = \alpha(IR)^2/Af$, where I is the current apply, R resistance and A is area) between different MTJs. We have not observed any dependence of the TMR and normalized noise on the junction area (A), which ranged between 100 µm² and 2500 µm². More details on the experimental setup may be found in Refs. [10, 11]

As the [110] interatomic distance in bulk MgO $a_{MgO}/\sqrt{2}$ = 0.298 nm is between these two former values, the misfit between the MgO and Fe-V layers varies with V concentration. It is well known that during a layer by layer growth process, the growing film is pseudomorphic to the substrate up to a critical thickness $h_c$ at which plastic relaxation occurs and dislocations nucleate. The lower is the misfit, the higher the critical thickness, and the lower the dislocations density after plastic relaxation. This means that the dislocations density in the MgO barrier is reduced when increasing the V content. As above $h_c$, the lattice parameter of the film increases with $h$ and tends to reach the bulk value, a good way to determine $h_c$ is to measure the average lattice parameter during the growth, which is possible by using electron diffraction. The average surface in-plane lattice parameter of MgO has thus been measured by RHEED during the growth on several Fe-V alloys. For this purpose, we monitored the distance between (220) and ($\bar{2}\bar{2}0$) diffraction rods, which is inversely proportional to the average surface parameter. As shown in Fig. 1 sudden increase of the average in-plane distance is observed in all cases. This behaviour corresponds to the appearance of the dislocations in the MgO film that changes the average lattice spacing, that is to $h_c$ [12]. The corresponding critical thickness $h_c$ is about 5 ML on Fe and substantially increases with $x$ up to 10 ML with 30% of V. Even though dislocations will be still present in our 9.5 ML thick MgO barriers for $x<0.3$, their density is reduced. Considering the *pessimistic* case of a fully relaxed film, the average distance $L$ between two dislocations in the MgO [100] or [010] in-plane directions would be $L(x) = a_{MgO}/2f(x)$ where the misfit $f(x) = \dfrac{a_{MgO}}{a(x)\sqrt{2}} - 1$. The dependence of $L$ with $x$ is likely to be stronger in our thin films stacking due to a partial relaxation.

Note that the thinner a stressed film is, the lower the required energy for nucleating a dislocation. Therefore, dislocations are easier to nucleate in a growing MgO film than in a completed one. The density of dislocations is thus mainly determined by the mismatch with the supporting (bottom) layer and to a lower extent by the top one. As a consequence, for an MTJ with two different electrodes, the density of dislocations depends on the stacking sequence. It is lower in an $Fe_{1-x}V_x$/MgO/Fe MTJ than in an Fe/MgO/$Fe_{1-x}V_x$ one. It should be noted that this variation of the dislocation density deduced from critical measurements was actually checked by HRTEM images analysis [12].

Figure 2(a) presents typical noise power spectra times area ($\alpha \times A$, for which Hooge factor is being analysed) when measured in the P state for the junctions with undoped, bottom doped or top doped electrodes. Clearly, V doping of the bottom Fe electrode tends to reduce the low frequency contribution to the noise, while V doping of the upper electrode tends to enhance the normalized 1/f noise. The inset, by expansion of the spectra up to 2000Hz, shows that noise for MTJ with bottom doped electrode may be accounted by the junction resistance in the conditions of direct electron tunnelling.

Figure 2(b) shows the variations of the averaged each set (x) at zero bias TMR and normalized 1/f noise (Hooge parameter) with V content. The error bars represent the dispersion of the corresponding values measured within each junction sets. In our convention, negative $x$ values correspond to a bottom $Fe_{1-x}V_x$ electrode, whereas positive $x$ values correspond to a top $Fe_{1-x}V_x$ electrode. In the latter case (top $Fe_{1-x}V_x$ electrode), the TMR is systematically lower than the one of standard Fe/MgO/Fe MTJs, reaching 185% at RT. It was shown elsewhere that this TMR drop is due to the reduced spin polarization of $\Delta$ states in Fe-V alloys [12]. In presence of a bottom Fe-V electrode and for $x<0.16$, we observe an increase of TMR from 185% to 207%, whereas the TMR decreases for larger V contents. These trends are similar to those observed with a thicker MgO barrier (12ML), the TMR reaching 240% at RT in the latter case [12] due to a better spin filtering when increasing the MgO barrier [17]. It was shown that the optimum TMR results from the competition between the reduction of the electrodes polarization (detrimental to the TMR) and the structural improvement of the barrier (beneficial to the TMR) [12]. Indeed, when used as the supporting *bottom* electrode, Fe-V alloys reduce the dislocations density and therefore the strain of the barrier and its roughness.

*Our main experimental finding* is the observation of a strong (nearly two orders of magnitude) decrease of the Hooge factor measured in the P state (i.e. defect related noise) with V doping of the *bottom* Fe electrode. We note that the minimum in the normalized nonmagnetic noise ($\alpha_P$) roughly coincides with the maximum TMR of about 207%. This corresponds to an enhancement of the signal to nonmagnetic noise ratio by more than 2 orders of magnitude in $Fe_{1-x}V_x$/MgO/Fe (0.08≤ $x$ ≤0.16) MTJs, in comparison with the reference Fe/MgO/Fe one. Although additional V doping further decreases the lattice mismatch at the Fe-V/MgO interface, contrary to expectations, the normalized noise values start to increase when $x$ exceeds 0.16. One possible reason could be the increased chemical disorder at the Fe-V/MgO interface, and the related spatial fluctuations of the potential.

In contrast, V doping of the upper electrode in Fe/MgO/Fe-V MTJs results in an increase of the nonmagnetic noise. Contrary to the case of a bottom Fe-V electrode, a *top* Fe-V one acts much less on the strain of the barrier because dislocations mainly nucleate during the growth of MgO. Here, the noise variations could reflect the increasing chemical disorder.

The normalized difference of the 1/f noise between AP and P states defined as ($\alpha_{AP}$ - $\alpha_P$)/$\alpha_P$ (see Figure 3) shows an interesting trend as a function of V alloying which may be important for applications of these MTJs. One observes that an enhanced TMR with V alloying is accompanied by a reduced relative noise in the AP state. We tentatively attribute this unexpected behaviour to Fe-V band hybridization. It is known that the minority, in contrast to majority spins, the Fe and V sites have an almost identical potential and can better hybridize [18]. This may result in an enhanced impurity charge screening in the Fe minority Δ bands. In other words, the V impurity charges may provide effective screening of scattering on relaxator-type defects for the Fe minority bands. As long as minority Δ bands control the conductance in the AP state [1,2], the Fe-V minority band hybridization may reduce the corresponding 1/f noise level.

**Figure captions**

Figure 1a – Variation of the average surface in-plane lattice parameter of MgO measured by RHEED during the growth on several Fe-V(001) surfaces.

Figure 2 (a) Voltage noise power spectral density times area measured on the junctions with undoped, bottom doped or top doped electrodes with bias of 200mV in the P state. The insert expands up to 2000Hz the power spectral density for the junction with bottom doped electrode. The green horizontal line marks the noise power times area expected level of $Fe_{92}V_{08}$/MgO/Fe with resistance of 160Ω and direct electron tunneling processes.
(b) Dependence of zero bias TMR and normalized noise (Hooge factor), averaged over each set, as a function of V content in the bottom (x < 0) and upper (x > 0) electrodes.

Figure 3 Normalized relative variation of the noise between antiparallel and parallel alignments, as determined from Fig 2(b) as a function of V content.

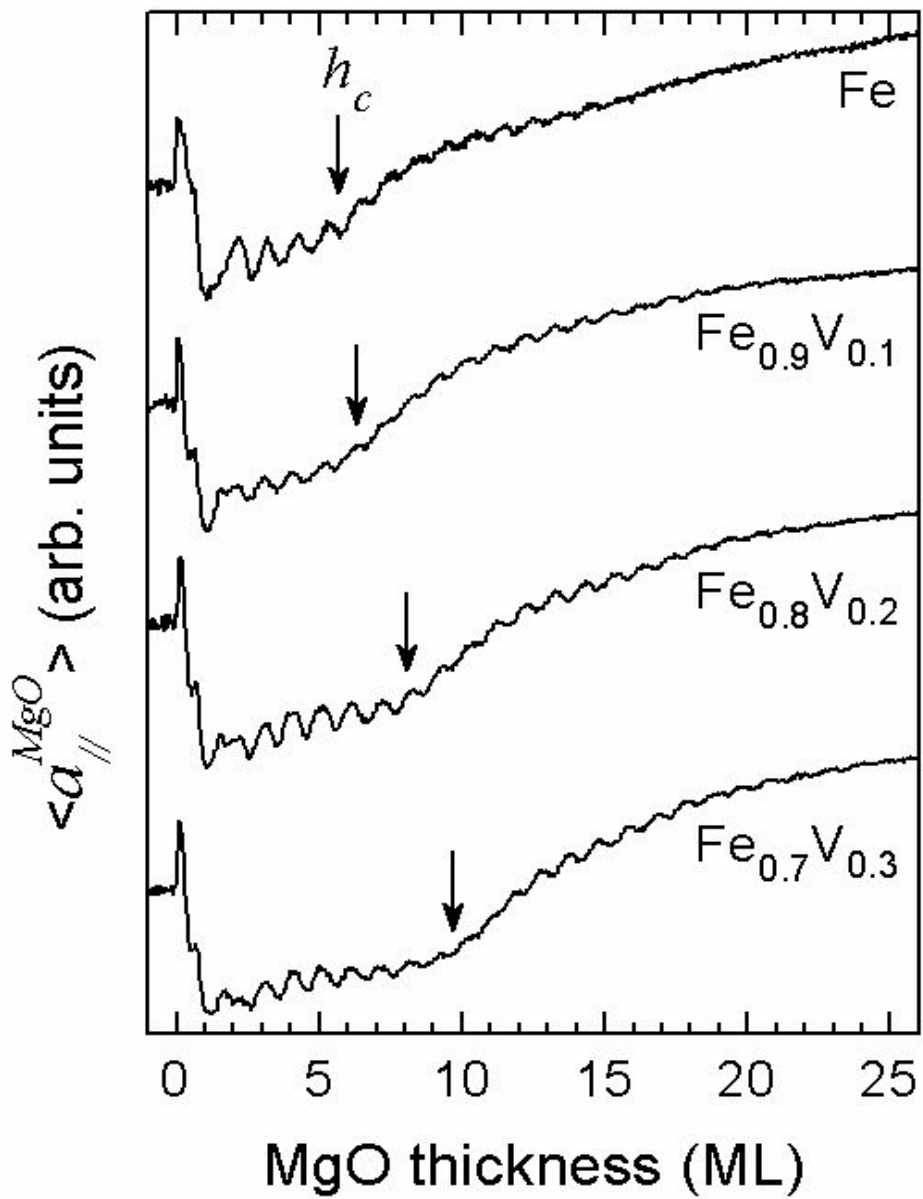

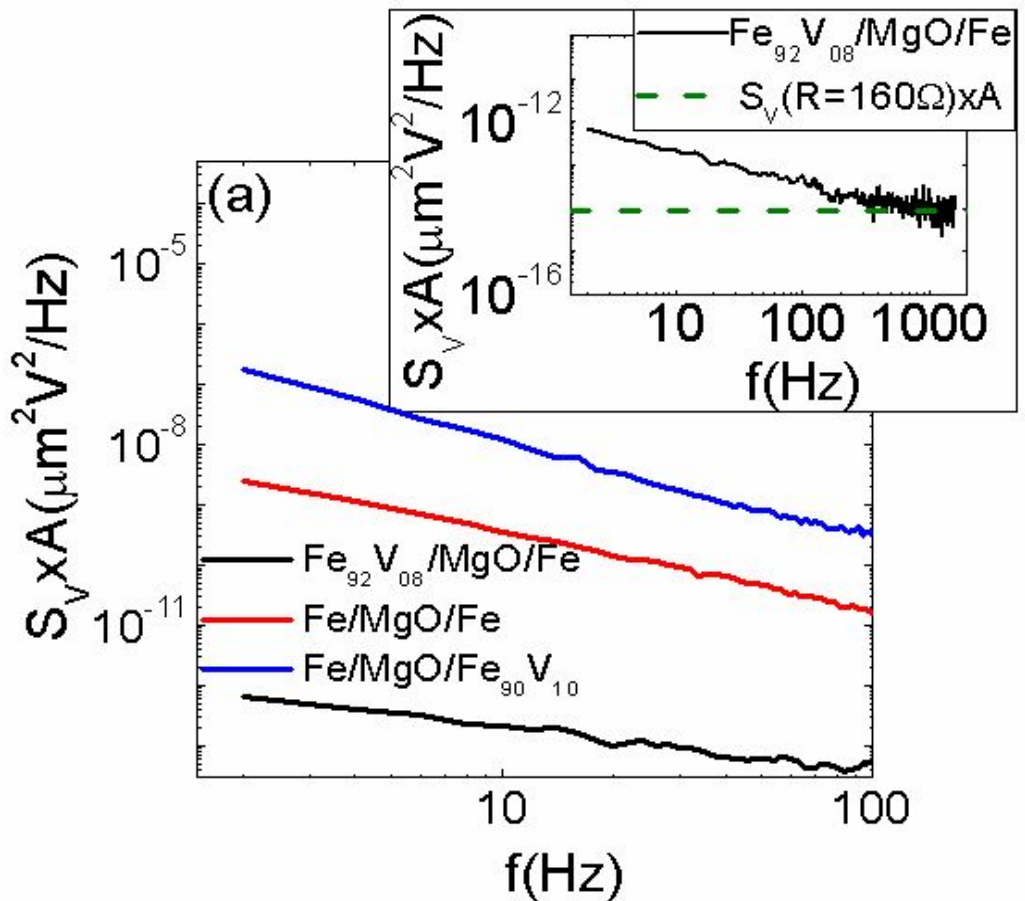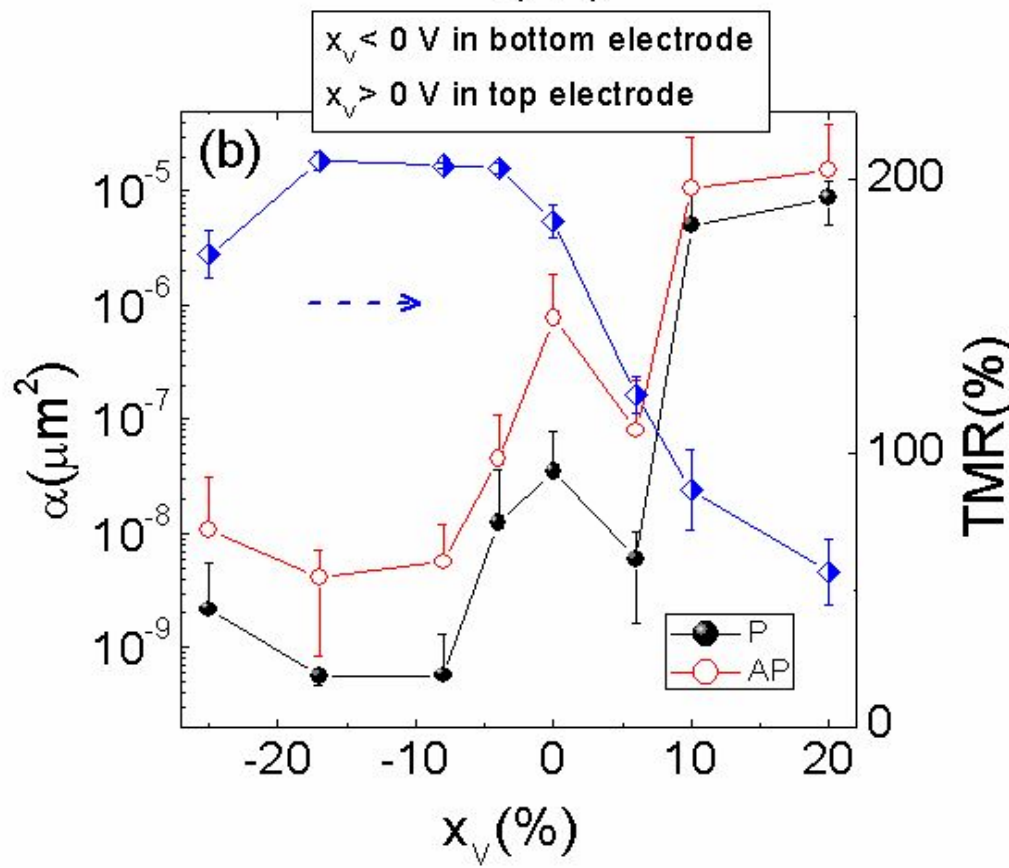

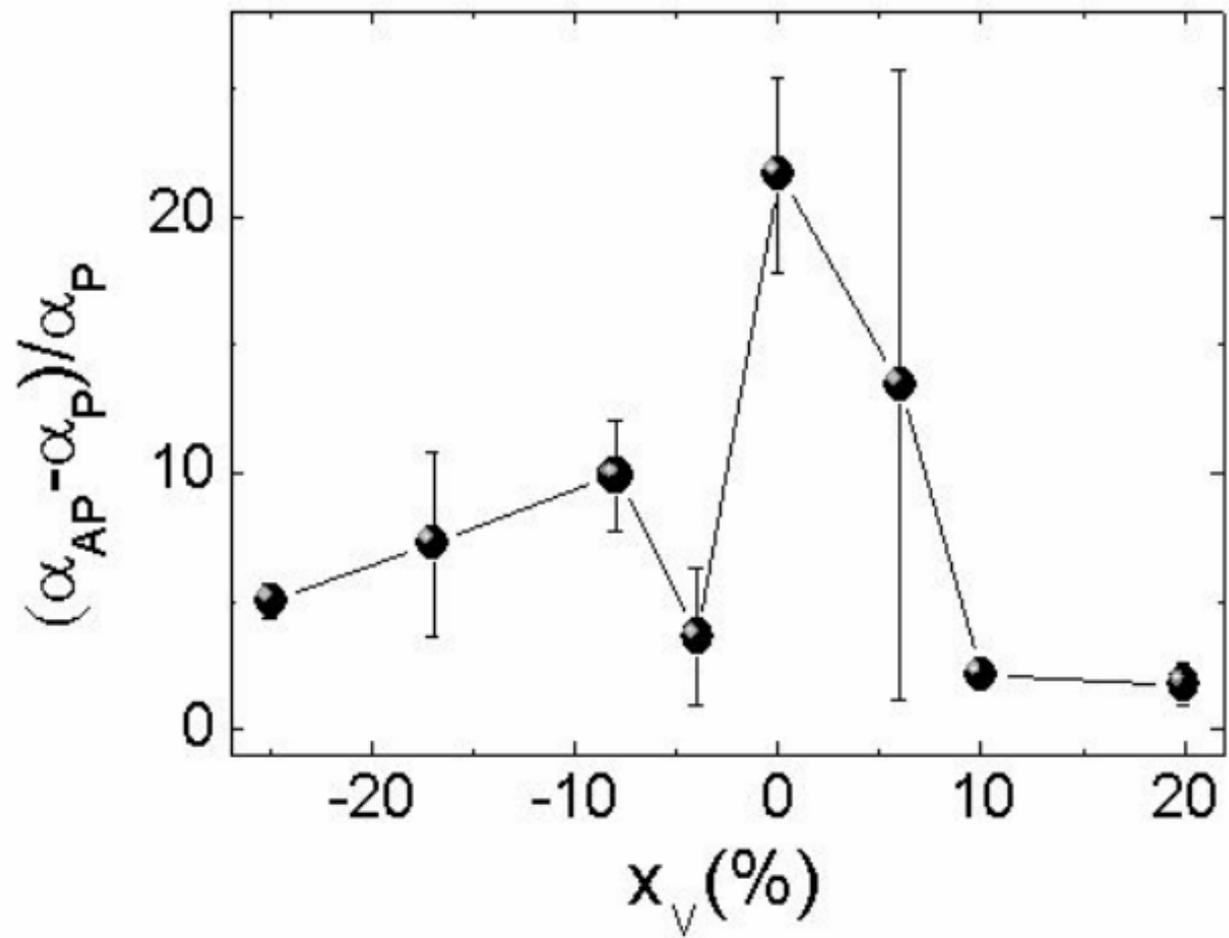